# BiFeO$_3$ nanocrystals for bio-imaging based on nonlinear optical harmonic generation

Sebastian Schwung,[1,6] Andrii Rogov,[2] Gareth Clarke,[3] Cécile Joulaud,[4] Thibaud Magouroux,[2] Davide Staedler,[5] Solène Passemard,[5] Thomas Jüstel,[6] Laurent Badie,[7] Christine Galez,[4] Jean Pierre Wolf,[2] Yuri Volkov,[3] Adriele Prina-Mello,[3] Sandrine Gerber-Lemaire,[5] Daniel Rytz,[1] Yannick Mugnier,[4,a)] Luigi Bonacina,[2,a)] Ronan Le Dantec[4]

[1]*FEE Gmbh, Struthstrasse 2, 55743 Idar-Oberstein, Germany*
[2]*GAP-Biophotonics, Université de Genève, 22 Chemin de Pinchat, 1211 Genève 4, Switzerland*
[3]*Department of Clinical Medicine, Trinity College Dublin, Ireland*
[4]*Univ. Savoie, SYMME, F-74000 Annecy, France*
[5]*Institute of Chemistry and Chemical Engineering, Ecole Polytechnique Fédérale de Lausanne, Batochime (Bat BCH), CH-1015 Lausanne, Switzerland*
[6]*Fachbereich Chemieingenieurwesen, Fachhochschule Münster, Stegerwaldstrasse 39, 48565 Steinfurt, Germany*
[7]*Institut Jean Lamour, UMR CNRS 7198, Faculté des Sciences et Techniques, Université de Lorraine, 54506 Vandoeuvre-lès-Nancy, France*

Abstract: Second Harmonic Generation (SHG) from BiFeO$_3$ nanocrystals is investigated for the first time to determine their potential as biomarkers for multiphoton imaging. Nanocrystals are produced by an auto-combustion method with TRIS as a fuel. Stable colloidal suspensions with mean particle diameters in the range 100-120 nm are then obtained after wet-milling and sonication steps. SHG properties are determined using two complementary experimental techniques, Hyper Rayleigh Scattering and nonlinear polarization microscopy. BiFeO$_3$ shows a very high second harmonic efficiency with an averaged $<d>$ coefficient of 79±12 pm/V. From the nonlinear polarization response of individual nanocrystals, relative values of the independent d$_{ij}$ coefficients are also determined and compared with recent theoretical and experimental studies. Additionally, the particles show a moderate magnetic response, which is attributed to γ-Fe$_2$O$_3$ impurities. A combination of high nonlinear optical efficiency and magnetic response within the same particle is of great interest for future bio-imaging and diagnostic applications.

[a)] Authors to whom correspondence should be addressed. Electronic mail: yannick.mugnier@univ-savoie.fr; Luigi.Bonacina@unige.ch

# I. INTRODUCTION

Harmonic nanoparticles (HNPs) are new types of exogenous contrast agents which have been recently proposed as biomarkers for multiphoton imaging.[1-6] They are based on noncentrosymmetric nanocrystals where efficient bulk second harmonic generation is allowed by the crystal symmetry. Due to the non-resonant nature of SHG, these biomarkers show very attractive properties for in vitro or in vivo imaging such as extreme photostability[4] and excitation wavelength tunability.[7] As underlined in recent reviews,[8,9] HNPs can complement usual fluorescent probes (i.e. dyes or quantum dots) for specific demanding applications like regenerative medicine.

From a material point of view, ideal HNPs must have high nonlinear optical (NLO) properties, low cytotoxicity but also easy processing for an effective control of the nanoparticles size, shape and surface properties. Up to now, various types of noncentrosymmetric nanoparticles have been produced, among which are numerous oxides such as $Fe(IO_3)_3$,[10] $LiNbO_3$,[11,12] $KNbO_3$,[13] KTP[14] and $BaTiO_3$[6,15] with typical diameters at about 100 nm or semiconductors like ZnO[16] and CdTe.[17] We showed in a recent comprehensive survey[18], that the harmonic conversion efficiency of most of these nanoparticles is close to the value for the corresponding bulk materials and thus, the responses of the different oxide nanoparticles are relatively similar. Moreover, we demonstrated that these nanoparticles have low cytotoxicity with the exception of ZnO particles.[18] Lastly, it was demonstrated that CdTe[17] quantum dots have noticeably higher intrinsic nonlinear efficiency but the known toxicity of these cadmium-based dots may appear as a limitation for in vivo applications.[19]

On the other hand, $BiFeO_3$ (BFO) is a widely investigated compound since it is, to date, the only known room temperature multiferroic material. Being extensively studied for its magnetoelectric behavior, BFO also has attractive ferroelectric properties with high spontaneous polarization.[20] Moreover, recent studies give evidences of a very promising nonlinear optical efficiency[21,22] which was found to be superior to the standard oxide materials previously introduced. Secondly, bulk BFO is antiferromagnetic but nanoparticles show weak ferromagnetism which was attributed to various possible causes: size effects,[23] extrinsic effects due to $\gamma$-$Fe_2O_3$ impurities[24] or oxygen vacancies.[25]

It is well-established that magnetic nanoparticles are essential for the biomedical field with key applications such as magnetic separation, Magnetic Resonance Imaging (MRI) or hyperthermia.[26] Accordingly, multifunctional BFO nanoparticles present a unique combination of potentially high nonlinear optical properties and magnetic characteristics which make them promising for the targeted bio-imaging applications and possible therapeutic applications.[27]

Here, we describe an extensive study on BFO nanoparticles preparation followed by nonlinear optical and magnetic characterization. BFO nanoparticles were produced by an original combustion synthesis with the aim of using a scalable production method and of lowering the annealing temperature. Indeed, to eliminate undesirable secondary phases during BFO preparation, the temperature must be high enough to form $BiFeO_3$[28] but at the same time it should stay low enough to prevent the thermal decomposition to $Bi_2Fe_4O_9$ and $Bi_{25}FeO_{39}$[29] or the partial reduction of $Fe^{3+}$ to $Fe^{2+}$[30]. In addition, a low temperature process preventing particle growth by Ostwald ripening is desirable, an aspect which is generally important for the synthesis of nanocrystals. After the preparation step, Hyper Rayleigh Scattering (HRS) measurements and SHG microscopy were used to determine the second order nonlinear optical properties of the particles in order to evaluate their potential as HNPs. Finally, vibrating sample magnetometry was used to measure their magnetic response.

## II. SYNTHESIS OF BFO NANOPARTICLES BY COMBUSTION METHOD

Different processes have been described for BFO preparation in the literature. Low process temperatures can be achieved by hydrothermal synthesis[31,32] also combined with microwave assisted heating[33]. Another method is the polyol-mediated synthesis in poly-alcohols like ethylene glycol or diethylene glycol.[23,34,35] However, these approaches remain undemonstrated for producing large amounts of material and a reduction of $Fe^{3+}$ by oxidation of the polyol was also observed.[36] On the other hand, widely used methods for the preparation of nanocrystalline BFO are sol-gel reactions like chelate assisted or Pechini reactions with citric acid[37] or tartaric acid.[38,39] As described in the original patent,[40] the Pechini method needs an annealing step to decompose the amorphous organic precursor and to promote crystallization.[39] A cross-linked network is indeed first produced where the metal ions are bonded to organic radicals by oxygen bonds.

With this in mind, the formation of a crystalline or partially crystalline precursor would appear to be a desirable improvement. For this purpose, the classical chelating agents can be substituted by the fuel 2-amino-2-hydroxymethyl-propane-1,3-diol (TRIS). Such combustion syntheses are known for oxide materials.[41] Briefly, this method is based on the highly exothermic redox reaction between an oxidizing agent such as metal nitrates and a reducing agent, i.e. the fuel.[41] During the ignition of the fuel, local temperatures can reach up to 2000 K.[42] In the case of BFO, sucrose-assisted synthesis has been described in the literature[43] where a stoichiometric amount of sucrose (fuel) was adapted to the metal ion concentration. The mixture was subsequently dried to a resin and auto-ignited on a magnetic stirrer. In another work on the auto-combustion synthesis of BFO, glycine[44] was used as a fuel.

The goal of the present work is to take advantage of the combustion method for the direct formation of a crystalline precursor of BFO following the first reaction step. With respect to standard sol-gel reactions, lower annealing temperatures can be expected to obtain phase pure

BFO, preventing Ostwald ripening and resulting in powder with smaller crystallites and better crystallinity.

All reagents were of analytical grade and were used without further purification or treatment. Iron(III) nitrate nonahydrate [Fe(NO3)3*9H2O], 99.5% purity (Merck) and bismuth nitrate pentahydrate [Bi(NO3)3*5H2O], ASC 98% (Alfa Aesar) were used as starting materials. The iron and bismuth contents were experimentally determined by complexometric titration. The precursor solutions of bismuth and iron are prepared by adding stoichiometric proportions (10 mmol) of the raw materials in diluted nitric acid under heating and stirring. Gradually, 18 g of 2-Amino-2-hydroxymethyl-propane-1,3-diol (TRIS) is added to the solution. The solvent is evaporated at 105 °C on a heating plate, forming a dried resin which ignites itself. The combusted resin is then ground in a mortar and calcined at temperatures between 400 and 600 °C for 1 hour to eliminate secondary phases. The as-obtained powder is wet milled on a rolling bench for up to 5 days to deagglomerate the primary particles in a gentle manner without destroying the crystallites. The block diagram of the general protocol is given in Fig. S1.[45]

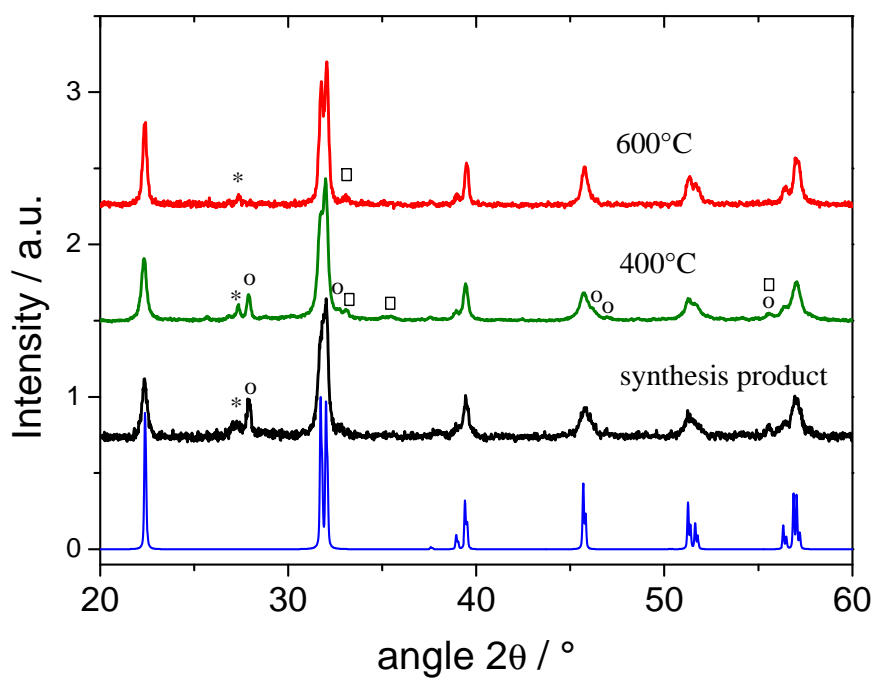

FIG 1. XRD patterns of BFO samples following the initial step of the combustion synthesis and after two subsequent annealings at 400°C and 600°C. Peaks denoted as *, o and □ belong to the parasitic phases $\gamma$-$Bi_2O_3$, $\beta$-$Bi_2O_3$ and $\alpha$-$Fe_2O_3$, respectively, while the others are in good agreement with the BFO reference profile (ICSD#15299) shown at the bottom of the graph.

The XRD pattern of the resulting synthesis product, obtained directly after the combustion step (i.e., without any annealing) is shown in Fig. 1. Interestingly, the $BiFeO_3$ phase is already predominant, with only a few minor peaks originating mainly from the parasitic phases $\beta$-$Bi_2O_3$ and $\gamma$-$Bi_2O_3$. After an annealing step at 400°C, those residual impurities are still present together with the $\alpha$-$Fe_2O_3$ phase after the decrease of the amorphous background. Phase purity is then improved with temperature increase to 600°C. Nonetheless, batch-to-batch variability was identified by the presence of differences in the impurity phases between samples. This is mainly associated with the timing and detailed procedure of the auto ignition which is difficult to control (see for instance samples BFO-HT1 and BFO-HT2 in Fig. S2).[45] After the combustion step, powder is wet milled on a rolling bench to reduce the particle size.

Representative image in Fig. 2 reveals nanoparticles without well-defined shape and typical dimension in the 100-250 nm range.

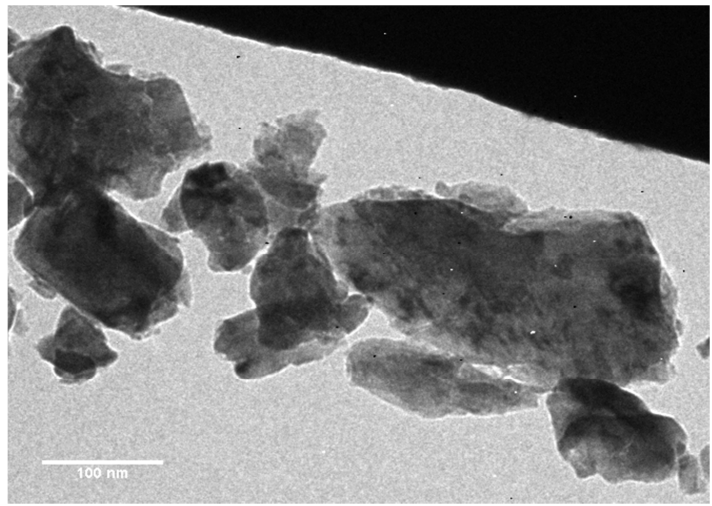

FIG 2. Representative TEM image of BFO particles annealed at 600°C.

## III. BFO COLLOIDAL SUSPENSION

Preparation of stable colloidal suspensions is needed for HRS studies and they are prepared using the following protocol: (i) Nanopowders are first dispersed into deionized water (pH=7) with a typical concentration of 0.4 mg/mL. (ii) Dispersions are exposed to ultrasonication for 25 min (Vibra Cell 75043, Bioblock Scientific) at a maximum input power of 750 W and a frequency of 20 kHz. A pulsed irradiation (1 s on and 4 s off) at room temperature was found to be optimal for the preparation of homogeneous dispersions.[46] (iii) Solutions are then left to settle for about one week to allow sedimentation of the larger particles and aggregates. (iv) The nanocrystal concentration at the end of the sedimentation period is estimated by weighing the residual fraction (assuming spherically shaped nanoparticles). It corresponds to 10-30% of the initial concentration.

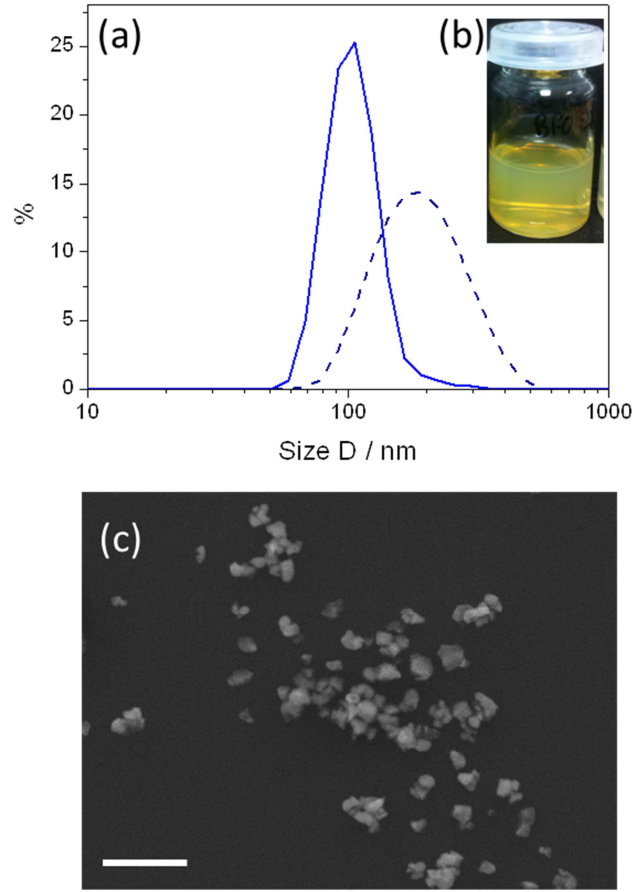

FIG. 3. (a) DLS size distribution by intensity (dash line) and by number (full line) and (b) suspension image. (c) SEM micrograph of the particles. Scale bar is 1μm.

Typical colloidal suspensions appear yellowish (Fig. 3b), due to partial absorption of BFO below 600 nm[47]. Suspensions are stable with values of zeta potential below -30 mV at pH 7. Mean particle sizes were estimated by Dynamic Light Scattering (DLS) at about 100-120 nm for most of the samples as also confirmed through SEM observations (Fig. 3a and 3c). Finally, it can be noticed that the particle size and shape polydispersity are reduced after this dispersion step, a necessary improvement for correct HRS measurements.

## IV. NONLINEAR OPTICAL PROPERTIES

BFO belongs to space group R3c and the SHG tensor can be described with four nonzero independent components: $d_{22}$, $d_{15}$, $d_{31}$ and $d_{33}$.[48] Very few studies have focused on the BFO optical second harmonic properties but, recently, the relative values of the $d_{ij}$ tensor coefficients were determined at 1064 nm from single crystal measurements.[49] In another study,[22] a careful determination of the absolute $d_{ij}$ coefficients was performed at 1550 nm in the case of both rhombohedral and strain-induced tetragonal thin films. Experimental results were found to be consistent with ab initio calculations.[21,50] Interestingly, the magnitude of the coefficients of BFO appears higher than the values known for commonly used nonlinear optical materials, making BFO a very good candidate for frequency conversion applications. In addition, a significant result of this study is that a large increase of the NLO response was observed for tetragonal films in agreement with other reported enhanced properties in such films.[51] In the present work, two different experimental techniques, namely HRS measurements and nonlinear polarization microscopy, were used to quantitatively assess the nonlinear optical response of BFO nanocrystals.

**A. Hyper Rayleigh Scattering**

HRS is an ensemble measurement, well-suited to readily determine the averaged nonlinear optical coefficients of nanocrystals. This technique was applied following the procedure previously detailed for the characterization of several noncentrosymmetric oxide nanocrystals such as $BaTiO_3$, ZnO, $LiNbO_3$ and $KNbO_3$.[18,52,53] HRS intensity results from the incoherent sum of the scattered SHG radiations emitted by each nanoparticle in suspension. The signal is thus proportional to the nanoparticles concentration, N, and can be expressed as:

$$I_{HRS} = GNT_{np} <\beta_{np}^2> I_\omega^2. \tag{1}$$

Here, G is an experimental constant and $T_{np}$ an internal field factor which is calculated according to the solvent and nanocrystal refractive indices (here $n_\omega$~2.76 and $n_{2\omega}$~3.20 for BFO[47]). $<\beta_{np}>$ is the nanoparticle effective hyperpolarizability from which the averaged SHG coefficient $<d>$ can be deduced if the nanoparticle volume $V_{np}$ is independently estimated:

$$<\beta_{np}> = <d> V_{np}. \tag{2}$$

This expression implicitly assumes that the measured SHG signal predominantly originates from the bulk contribution thus omitting any possible surface contribution. For oxide nanomaterials, this type of behavior was already demonstrated for several nanoparticles in the 100 nm range[53] as well as more recently for smaller nanocrystals.[11,15] In the above formulae, brackets indicate isotropic orientational averaging. The relation between $<d>$ and the tensor components of BFO can be written, in our experimental configuration, as:

$$<d^2> = \tfrac{6}{35}d_{33}^2 + \tfrac{32}{105}d_{31}^2 + \tfrac{8}{21}d_{22}^2 + \tfrac{44}{105}d_{15}^2 + \tfrac{4}{21}d_{31}d_{33} + \tfrac{4}{35}d_{15}d_{33} + \tfrac{16}{105}d_{31}d_{15} \ . \quad (3)$$

Experimentally, the nanoparticle suspension was excited with a vertically polarized YAG laser ($\lambda_\omega$=1064 nm) and the scattered, unpolarized, SHG signal was detected at 90 degrees from the laser beam axis with a photomultiplier. Calibration of the experimental set-up was previously performed with a molecular solution of para-nitroaniline as an external reference. For each sample, the HRS intensity was then measured according to the nanoparticle concentration. Quantitative values for $<\beta_{np}>$ and $<d>$ were finally derived from the nanoparticle concentration and size independently obtained through weighing and DLS measurements, respectively.[53]

Fig. 4 shows a representative plot of the HRS intensity as a function of the BFO nanoparticle concentration. One can observe a deviation from the expected linear behavior for the highest concentrations. This is related to the suspension absorption mostly at the second harmonic frequency.[53] There is however a clear linear tendency for low nanoparticle concentrations, allowing the determination of the experimental slope which is used to calculate $<\beta_{np}>$ and $<d>$.

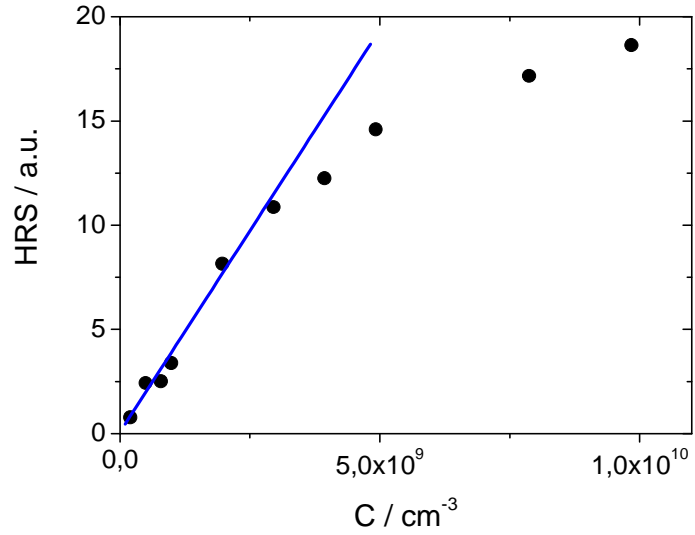

FIG 4. HRS intensity as a function of the concentration of nanoparticles in suspension.

Determination of these experimental parameters has been performed on four independent samples resulting from the different BFO batches (Table SI).[45] The effective hyperpolarizability is directly proportional to the nanoparticle volume (Eq. 2) and thus varies from sample to sample. For instance, a value of $<\beta_{np}> = 1.2.10^{-22}$ esu was found for a 110 nm particle mean size. On the contrary, the averaged coefficient $<d>$ is an intrinsic physical property of the nanomaterial itself which is not affected by the nanocrystal size. Experimental values range from 67 to 90 pm/V, depending on the sample, with a mean value of 79 pm/V. This variability is in line with the experimental reproducibility of the overall measurement process (estimated at ±15%) and is mainly related to the difficulty in precisely determining the particle size from DLS, as discussed in ref [53]. More importantly, there is no clear correlation between the type and concentration of impurities and the experimental $<d>$ value, as long as BFO is the dominant phase in the different synthetized powders (Fig. S2 and Table SI).[45] This finding is in agreement with the hypothesis of a predominant bulk SHG contribution related to the noncentrosymmetric structure of the material. The "actual

BFO nanocrystal volume" for each particle is indeed poorly affected, at first order, by the presence of minor impurity phases.

Finally, this first quantitative assessment of the NLO response of BFO nanocrystals is found to be in relatively good agreement with the $d_{ij}$ coefficients extracted from ab initio calculations.[21] At 1064 nm, these theoretical coefficients lead to a $<d>$ value of 45.4 pm/V (Table SII).[45] Valuable comparative information was also gained when comparing the NLO response of BFO with previously characterized nanomaterials under the same experimental HRS setup. Interestingly, it was found for instance $<d>$ = 7.4 pm/V and $<d>$ = 5.5 pm/V for $LiNbO_3$ and $BaTiO_3$ nanocrystals, respectively.[53] A very good nonlinear optical efficiency is therefore demonstrated for BFO which is typically one order of magnitude higher than for standard NLO materials.

**B. Nonlinear polarization microscopy**

The SHG images were acquired using an inverted laser scanning microscope (Nikon TE300), equipped with a 0.6NA 40×objective. The position of the sample was controlled by an XYZ piezo-scanner with 2 nm resolution. The excitation source was a mode locked Ti:Sapphire oscillator (Synergy 20, Femtolasers), providing 20 fs pulses at 80 MHz repetition rate. The incident light polarization was selected by a rotating half wave plate, while SHG signal was epi-detected using the same objective. A combination of a 650 nm blocking edge short pass filter (AHF, F37-650) and a bandpass filter (40 nm bandwidth at 400 nm) ensured an efficient rejection of the 800 nm scatter. After being analyzed by a Glan–Taylor polarization cube, the signal was measured by a photomultiplier tube and processed by a lock-in amplifier.

At first, an SHG raster scan image of a substrate with isolated nanoparticles was acquired. Then the laser beam was focused onto selected crystals. During this operation the SHG signal was maximized by adjusting the microscope piezo-scanner displacements. The polarization response was acquired by measuring the SHG signal as a function of the polarization angle of the incident light γ and of the detection polarization, setting the analyzer along the *X* or *Y* direction (definitions in Fig. 5).

The experimental SHG polarization response was then fitted according to the following assumptions[1]: the fundamental beam is assumed to be at normal incidence, with only two in-plane components $E_x^\omega = E_\omega \cos\gamma$ and $E_y^\omega = E_\omega \sin\gamma$; the collected SHG intensity in the *X* direction $I_x$ (or equivalently, $I_y$, in the Y direction) is then assumed to be proportional to $P_x^2$ ($P_y^2$), where *P* is the nonlinear polarization given by:

$$P_i^{2\omega} = \varepsilon_0 \sum_{jk} d_{ijk} E_j^\omega E_k^\omega . \qquad (4)$$

The crystal nonlinear tensor $d_{ijk}$ is given here in the laboratory reference frame and can be derived from the $d_{mnl}$ tensor expressed in the crystal frame by:

$$d_{ijk} = \sum_{mnl} d_{mnl} S_{im} S_{jn} S_{kl} , \qquad (5)$$

where *S*<sub>*im*</sub> are the components of the rotation matrix between the laboratory and crystal axes depending on the Euler angles φ, θ, and ψ. Knowing $d_{mnl}$ elements, four free parameters remain to be adjusted for the fit: the Euler angles and the experimental factor *K* which accounts for the incident laser intensity, the collection and the detection efficiency and the crystal size. For the data analysis, the polarization response of the dichroic mirror and the objective was also taken into account.

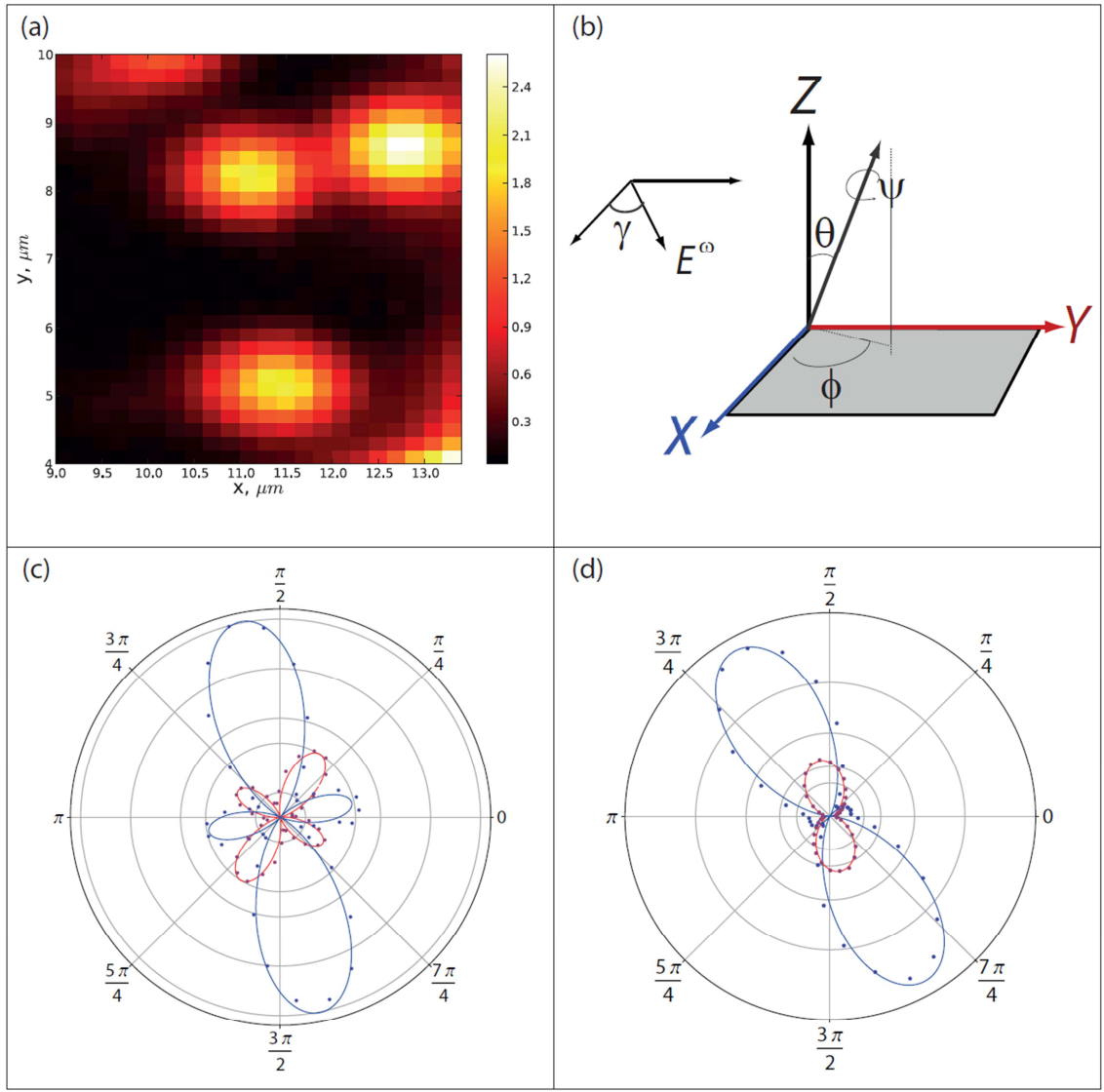

FIG. 5. (a) 2D SHG image of BiFeO$_3$ nanocrystals. (b) Definition of the angles. The crystal orientation is expressed in the laboratory frame X,Y,Z by the Euler angles φ, θ, and ψ. The angle γ denotes the polarization of incident laser light on the sample plane. (c,d) Polarization emission of BiFeO$_3$ nanocrystals. Each signal is analyzed along two orthogonal directions: X (blue color) and Y (red color). Dots correspond to experimental data, while solid lines represent the best fit.

Currently in the literature a few different sets of relative $d_{ij}$ values for BFO are present, as reported in Table SII.[45] The first set *(set 1)* was derived from BFO single crystal measurements[49] at 1064 nm, the second *(set 2)*[47] and third *(set 3)*[22] were obtained from BFO thin film measurements at 800 nm and 1550 nm, respectively. A fourth set *(set 4)* can also be obtained from ab initio simulation.[21] Although Haislmaier et al.[22] correctly pointed out that the set 2 of $d_{ij}$ values does not account for absorption at 400 nm, this set, which was derived for 800 nm excitation, is the only one which allowed us to fit our experimental results as shown in Fig. 5. Two representative results are shown in panel 5c and 5d, as results of the experimental fitting indicated above, and the same fitting procedure yielded similar agreement on all the particles in the sample except for large agglomerates.[1] It can be noticed that only the relative $d_{ij}$ coefficients are involved in the fitting procedure. We thus find that the relative $d_{ij}$ values of *set 2* correctly fit our experimental response. However it is to be pointed out that the averaged <d> coefficient computed from the individual tensor components given in [47] exceeds by two orders of magnitude our estimation based on HRS measurements (Table SII).[45]

## V. MAGNETIC PROPERTIES

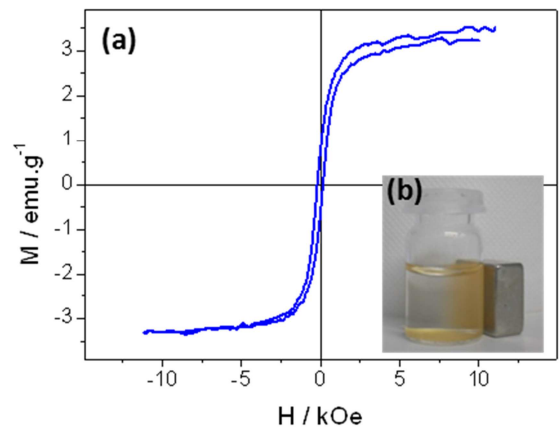

FIG. 6. (a) Room temperature magnetic hysteresis loops of BFO sample BFO-LT2. (b) Magnetic separation of BFO particles in water suspension.

Room temperature magnetic response, measured by vibrating sample magnetometry, for a BFO sample annealed at 350°C is shown in Fig. 6a (sample BFO-LT2 in Table SI).[45] A clear ferromagnetic behavior is measured with a relatively high saturation magnetization at 3.5 emu.g$^{-1}$. This can be related to the presence of impurities.[24,44] Pure BFO nanoparticles only display weak ferromagnetism, with saturation magnetization of about 1 emu.g$^{-1}$, when the particle size is below the period of the spiral spin structure at 62 nm.[23,54] On the other hand, stronger magnetic responses were already reported and traced to iron oxide impurities (it is worth noting that this behavior has already been seen in samples prepared by combustion synthesis).[44] This is in agreement with the XRD analysis presented in this work that clearly shows the presence of the γ-Fe$_2$O$_3$ phase in the sample, as shown in Fig. S2.[45]

Moreover, it should be pointed out that the SHG properties of the BFO particles are not affected by the presence of small quantities of impurity phases. For instance, the nonlinear $<d>$ coefficient of the BFO-LT2 sample is 84 pm/V, in agreement with other samples in this study. Thus, it shows that careful synthesis optimization can promote both second harmonic efficiency and high magnetic properties. This latter is useful for magnetic separation, for instance, as shown in Fig. 6b. Therefore, the multimodal potential of the particles presented in this work could be advantageously applied for specific biomedical applications.

## VI. CONCLUSION

BFO nanoparticles were prepared by the combustion method and their SHG properties were then determined through HRS and nonlinear polarization microscopy. The high SHG efficiency, with a $<d>$ coefficient measured at 79±12 pm/V, confirms that BFO is a promising material for frequency conversion applications. We additionally demonstrated that the nanoparticles SHG properties are not quenched when small quantities of impurities are present. This is due to the bulk nature of the second harmonic process occurring within BFO particles. Moreover, moderate ferromagnetic response was observed for nanoparticles containing residual γ-Fe$_2$O$_3$ impurities, allowing the particles to be magnetically separated in solution. BFO nanoparticles are therefore new promising HNPs combining a very high SHG brightness and a moderate magnetic response suitable for a wide range of bio-imaging and diagnostic applications.

**ACKNOWLEDGEMENTS**

This research was partially supported by a European FP7 Research Project NAMDIATREAM (NMP4-LA-2010- 246479, http://www.namdiatream.eu).
The authors thank Michel Moret for the help with the design of mechanical components of the microscope involved in nonlinear microscopy measurements.

# Supplemental material:

## 1. Block diagram of the synthesis

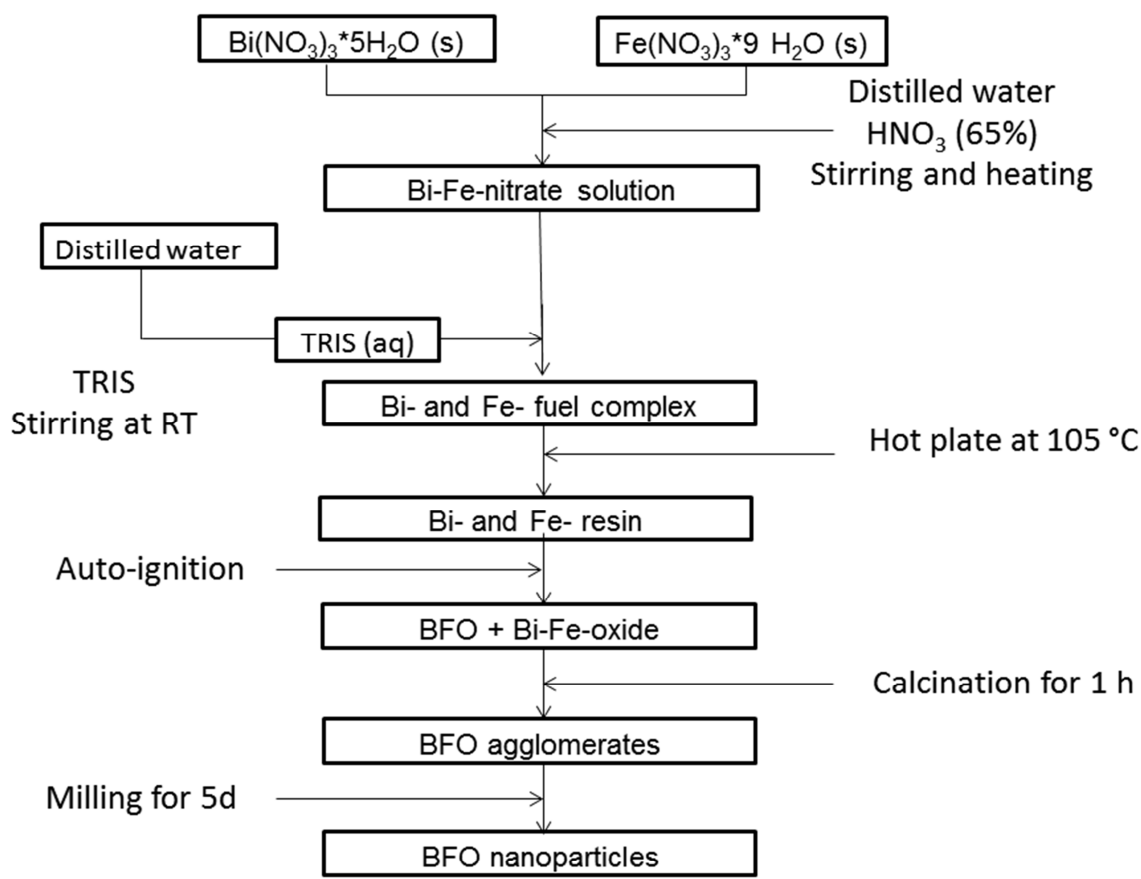

FIG S1. Block diagram describing the main steps of the combustion method used for BFO synthesis.

## 2. BFO samples used for optical and/or magnetic characterizations

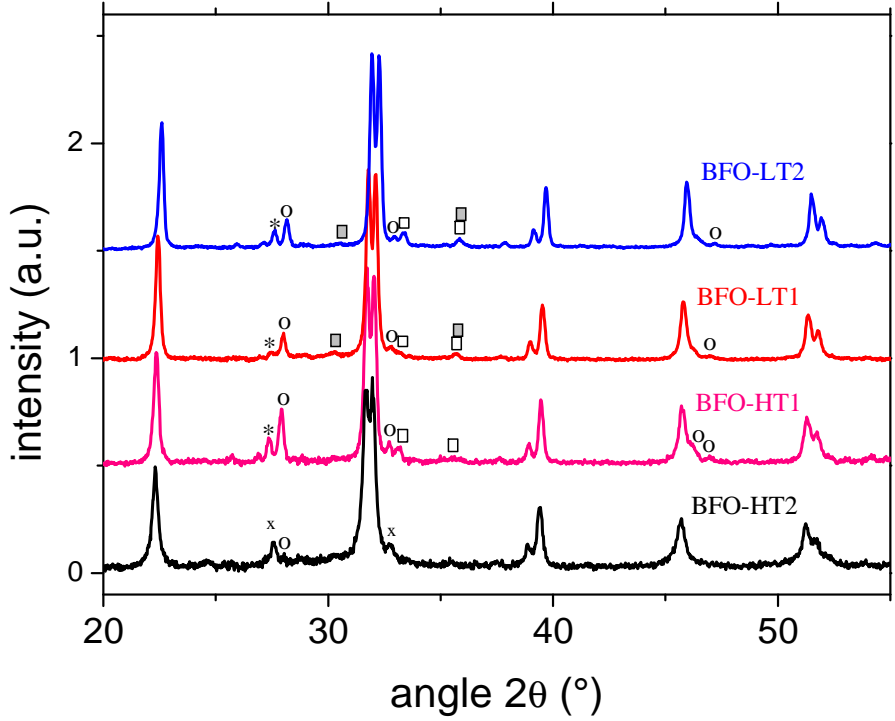

FIG S2. XRD patterns of BFO samples used for optical and/or magnetic characterizations. Peaks denoted as *, o, x belong to the Bi-rich phases $\gamma$-$Bi_2O_3$, $\beta$-$Bi_2O_3$ and $Bi_{25}FeO_{39}$. Peaks denoted as □, ■ belong to iron oxide phases $\alpha$-$Fe_2O_3$ and $\gamma$-$Fe_2O_3$. For low annealing temperature of 350°C, the ferromagnetic $\gamma$-$Fe_2O_3$ phase is clearly observed.

Table SI. Combustion synthesis parameters, experimental $<d>$ coefficient and saturation magnetization Ms for four BFO samples.

| Sample name | Annealing parameters | $<d>$ [pm/V] | $M_s$ [emu.g-1] |
|---|---|---|---|
| **BFO-HT1** | 600 °C, 1 h | 90 | -a) |
| **BFO-HT2** | 600 °C, 1 h | 76 | -a) |
| **BFO-LT1** | 350 °C, 1 h | 67 | 4.0 |
| **BFO-LT2** | 350 °C, 1 h | 84 | 3.5 |

a) Magnetic data are not available for this sample

## 3. Literature values of BFO nonlinear coefficients

Table SII. Nonlinear coefficients $d_{ij}$ (pm/V) values from the literature

| Nonlinear coefficients | (Set 1) Single crystal[1] 1064 nm[a] | (Set 2) Thin film[2] 800 nm | (Set 3) Thin film[3] 1550 nm | (Set 4) Ab initio[4] 800 nm | Ab initio[4] 1064 nm |
|---|---|---|---|---|---|
| $d_{22}$ | -0,028 | 298,4 | 18,7 | -33,2 | 54,0 |
| $d_{15}$ | 15 | 59,7 | -0,9 | -3,2 | -4,3 |
| $d_{31}$ | 1 | 104,4 | -8,5 | 60,4 | -21,1 |
| $d_{33}$ | -2,3 | -3401 | -15,1 | 50,3 | -55,9 |
| $<d>$[b] | - | 1389,9 | 14,9 | 50,0 | 45,4 |

[a] relative values respect to d31; [b] calculation using Eq. (3)

## References

[1] H. Yokota, R. Haumont, J.-M. Kiat, H. Matsuura, and Y. Uesu, Appl. Phys. Lett. **95**, 082904 (2009).

[2] A. Kumar, R.C. Rai, N.J. Podraza, S. Denev, M. Ramirez, Y.-H. Chu, L.W. Martin, J. Ihlefeld, T. Heeg, J. Schubert, D.G. Schlom, J. Orenstein, R. Ramesh, R.W. Collins, J.L. Musfeldt, and V. Gopalan, Appl. Phys. Lett. **92**, 121915 (2008).

[3] R.C. Haislmaier, N.J. Podraza, S. Denev, A. Melville, D.G. Schlom, and V. Gopalan, Appl. Phys. Lett. **103**, 031906 (2013).

[4] S. Ju, T.-Y. Cai, and G.-Y. Guo, J. Chem. Phys. **130**, 214708 (2009).